\documentclass{osa-article}

\journal{osac}
\articletype{Research Article}
\usepackage{amsmath}
\usepackage{braket}
\usepackage{mathtools}

\DeclarePairedDelimiterX{\abs}[1]{\lvert}{\rvert}{\ifblank{#1}{{}\cdot{}}{#1}}
\DeclarePairedDelimiterX{\norm}[1]{\lVert}{\rVert}{\ifblank{#1}{{}\cdot{}}{#1}}

\newcommand{\tens}[1]{%
  \mathbin{\mathop{\otimes}\limits_{#1}}%
}

\begin{document}

\title{Utilizing a Fully Optical and Reconfigurable PUF as a Quantum Authentication Mechanism}

\author{H Shelton Jacinto,\authormark{1,2,*} A. Matthew Smith,\authormark{1} and Nader I. Rafla\authormark{2}}

\address{\authormark{1}Quantum Information Science, Air Force Research Laboratory, Rome, NY 13440\\
\authormark{2}Department of Electrical and Computer Engineering, Boise State University, Boise, ID 83725}

\email{\authormark{*}sheltonjacinto@u.boisestate.edu}

%% To be edited by editor
% \dates{Compiled \today}

%\ociscodes{(140.3490) Lasers, distributed feedback; (060.2420) Fibers, polarization-maintaining;(060.3735) Fiber Bragg gratings.}

%% To be edited by editor
% \doi{\url{http://dx.doi.org/10.1364/XX.XX.XXXXXX}}

\begin{abstract}
In this work the novel usage of a physically unclonable function composed of a network of Mach-Zehnder interferometers for authentication tasks is described. The physically unclonable function hardware is completely reconfigurable, allowing for a large number of seemingly independent devices to be utilized, thus imitating a large array of single-response physically unclonable functions. It is proposed that any reconfigurable array of Mach-Zehnder interferometers can be used as an authentication mechanism, not only for physical objects, but for information transmitted both classically and quantumly. The proposed use-case for a fully-optical physically unclonable function, designed with reconfigurable hardware, is to authenticate messages between a trusted and possibly untrusted party; verifying that the messages received are generated by the holder of the authentic device.
\end{abstract}

%\maketitle

\section{Introduction}\label{sec:intro}

Authentication is an important part of any communication protocol, as it provides a method of guaranteeing the trustworthiness of users and other components in a network, such as keys and messages. Identity authentication is generally the first method considered, as this primitive provides protection against an eavesdropper, Eve, pretending to be a legitimate user \cite{bellare1993entity}. It is important to make any protocol resistant to an eavesdropper, such that the transmitted messages are known to be from authenticated users. In any authentication scheme the receiver verifies the creator and sender of information, where in two-party identity authentication specifically, a communicating party tries to prove (\textit{i.e.} the \textit{prover}) their identity and the other verifies (\textit{i.e.} the \textit{verifier}) the provided identity before trusting the data.

An identity authentication scheme works where a sender pre-registers confidential information regarding his or her identity, often in a form a key or per-user string, in a database held by the receiver, prior to any communication occurring. When communication is initiated, an identity authentication event happens where the receiver can receive a set of confidential information from the sender and verify it against the previously-registered information in the database. 

Other versions of authentication, specifically message authentication, work where the sender and receiver have a secure channel that may be in an untrusted environment. Within the untrusted environment there is the potential for an eavesdropper to intercept and manipulate messages, or pose as either the sender or receiver, in what is commonly known as a `man-in-the-middle' (MITM) attack. The message authentication techniques help to overcome the falsification of identity when sending messages by appending, or applying, hardware-specific user information to the message between a prover and verifier. The proposed work here represents a model for a quantum-enhanced message authentication scheme built around an integrated optical-hardware-based physically unclonable function (PUF).

The integrated photonic hardware used in this work is known as a quantum photonic processor (QPP) and was developed in collaboration between the Air Force Research Lab (AFRL) and the Quantum Photonics Laboratory at MIT under the direction of D. Englund \cite{englund17,harris18, Carolan15}. The form of optical PUF utilized differs from previous optical PUFs, such as that of Grubel \cite{Grubel:17} or Tarik \cite{Tarik20}, in that our device is a mesh-network of Mach-Zehnder interferometers (MZIs) rather than a single chaotic resonator. Other optical PUFs, including those based on bulk materials, have been developed such as multi-mode fibers and other scatting media, along with authentication methods \cite{Goorden:14,Yao20,Pappu02}. A detailed study of the theory of classical and quantum PUFs is available in \cite{Gianfelici2020}.

\section{Previous Work}\label{sec:prev_work}

\begin{figure}[htb]
    \centering
    \includegraphics[width=0.65\columnwidth,keepaspectratio]{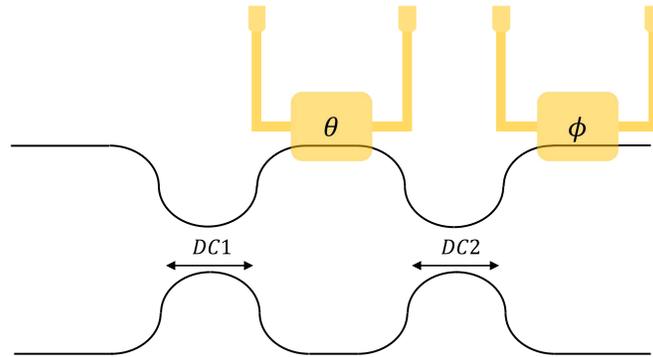}
    \caption{\textbf{Integrated design} of the Mach-Zehnder interferometers used in the quantum photonic processor showing the waveguides being acted on by computer-controlled phase modulators. The phase modulators adjust the internal ($\theta$) and external ($\phi$) phases seen at the output ports of the device with two 50:50 beamsplitters implemented as directional couplers, DC1 and DC2.}
    \label{fig:mzi_basic}
\end{figure}

In previous work by A. Smith and H. Jacinto, a quantum photonics processor, implemented as a photonic integrated circuit (PIC), was employed as a physically unclonable function \cite{Smith20}. The fully-optical PUF is designed as a silicon-on-insulator (SOI) integrated optical chip with electronic control. The optical interferometric device consists of 88 2x2 MZIs, schematically shown in Figure~\ref{fig:mzi_basic}.

\begin{figure}[htbp]
    \centering
    \includegraphics[width=0.70\textwidth,keepaspectratio]{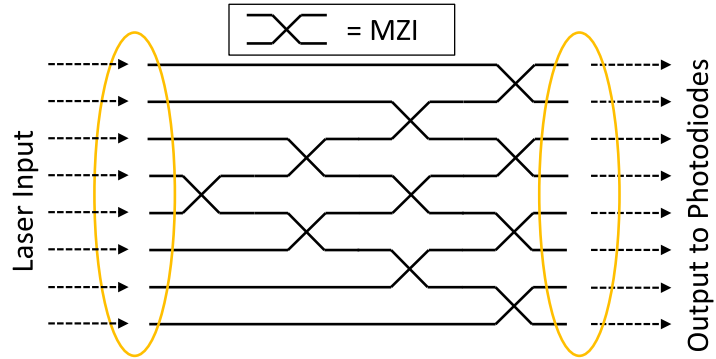}
    \caption{\textbf{Simplified MZI architecture} depicting the possible inputs that can be used and output readout via photodetectors. Advanced pumping schemes can be realized, utilizing external feedback, or where multiple inputs are being pumped by laser-light simultaneously. The MZI symbol at top is represented by a `cross' where each MZI is composed similarly to the one shown in Figure~\ref{fig:mzi_basic}.}
    \label{fig:mzi_puf_simple}
\end{figure}

The MZIs are connected in a triangular nearest-neighbor configuration, as shown in Figure~\ref{fig:mzi_puf_simple}. Each MZI is thermally tuned by integrated resistive heating elements on the upper waveguide, with the MZI applying an ideal $2\times2$ unitary transformation shown in Equation~\ref{eq:mzi_xfrfn}.

\begin{align} \label{eq:mzi_xfrfn}
	U_{MZI}(\theta, \phi) = &\dfrac{1}{2}
	\begin{pmatrix}
		e^{j\phi} & 0\\
		0 & 1
	\end{pmatrix}
	\begin{pmatrix}
		1 & j \\
		j & 1
	\end{pmatrix}
	\begin{pmatrix}
		e^{j\theta} & 0\\
		0 & 1
	\end{pmatrix}
	\begin{pmatrix}
		1 & j\\
		j & 1
	\end{pmatrix} 
	\nonumber\\= &je^{\frac{j\theta}{2}}
	\begin{pmatrix}
		e^{j\phi}\text{sin}\big(\frac{\theta}{2}\big) & e^{j\phi}\text{cos}\big(\frac{\theta}{2}\big)\vspace{3pt}\\
		\text{cos}\big(\frac{\theta}{2}\big) & -\text{sin}\big(\frac{\theta}{2}\big)
	\end{pmatrix}
\end{align}
Each MZI consists of two integrated phase shifters: One phase shifter between the two beam splitters, and a second phase shifter on one output leg. These four components are the four $2\times 2$ unitary matrices multiplied in sequence in Equation~\ref{eq:mzi_xfrfn}.The unitary transformation equation includes two variables, $\theta$ and $\phi$, which map to the internal phase setting and output phase offset, respectively, for each MZI.

\begin{figure*}[htp]
    \centering
    \includegraphics[width=1.0\textwidth,keepaspectratio]{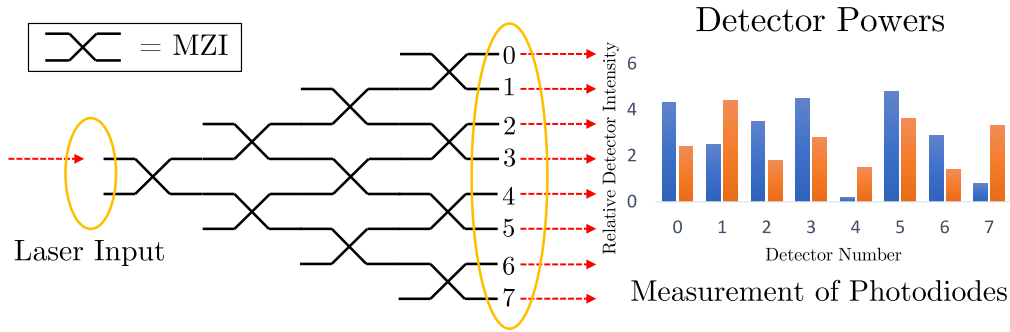}
    \caption{\textbf{Optical PUF output profile} from two example challenges applied to the (simplified) device in the form of MZI settings, with the right graph showing the resulting profile from the photodetector's response in terms of relative intensity. The blue and orange bars represent possible responses to a predefined set of two challenges.}
    \label{fig:mzi_puf_profile}
\end{figure*}

In this architecture's foundational application as a PUF \cite{Smith20}, the distinguishablility and uniqueness of the optical design were demonstrated under classical inputs. The metrics chosen by \cite{Smith20,Mesaritakis18} were the Euclidean distance and $\ell^2$-norm of the $N$ number of outputs, from a single laser-light input, depicted as $N=8$ in Figure~\ref{fig:mzi_puf_profile}. In addition, the inter- and intra-device Hamming distances, ${HD}_{inter}$ and ${HD}_{intra}$, modified as the loose Hamming distance, were used as a measure of relative distinguishability between groups of PUF settings \cite{Smith20}. The work in \cite{Smith20} defines the quality of the PUF and gives an estimate of the number of reconfigurable settings based on the physical architecture of the device; a value of $6.85 \times 10^{35}$ challenge-response pairs (CRPs) for the full device. The same work defines a uniqueness constraint to determine the number of fully-independent CRPs that can be utilized. Different architectures will change the number of viable CRPs. This work assumes that the large device from \cite{Smith20} is being used for all authentication described.

%\textcolor{blue}{
%Optical PUFs, such as speckle patterns \cite{Pappu02} and multilmode fiber \cite{Mesaritakis18} often require nontrivial bulk optical %components and alignment.  Our proposed solution avoids such hard-to-scale elements and can be tightly integrated with standard CMOS %components as well as the growing field of quantum optical devices. Previous PUFs also lack the rapid reconfigurability and %repeatability aspects discussed here and can be seen in greater detail in \cite{Smith20}.  }

Alternate forms of optical PUFs often require non-trivial bulk optical components, and alignment, to observe optical patterns originating from things such as micro-structured tokens generating speckle patterns \cite{Pappu02} and imperfections in multi-mode fibers \cite{Mesaritakis18}. Our proposed solution utilizes a PUF which avoids hard-to-scale elements, can be tightly integrated with standard CMOS components and processes, and effortlessly operates with devices from the growing field of integrated quantum optics. Of special note is our PUF's additional operating mode; where rapid reconfigurability and repeatability aspects allow for our device to operate as multiple PUFs. Many of the operating aspects of our PUF, which this work relies on, may be seen in greater detail in \cite{Smith20}.

\subsection{Security of Proposed Architecture Against Numerical Modeling Attacks}
Compared with traditional CMOS-based PUFs and other optical PUFs, the device utilized in \cite{Smith20} is unique in its resistance to numerical model-based attacks. The resistance comes from the numerical complexity present in the device to be used in this work, where the state-space for a CMOS-based PUF is comparatively small. From \cite{Smith20} Section IV.A, it can be seen that for a device similar to Figure~\ref{fig:mzi_puf_simple} with 10 MZIs, there exists $1.19 \times 10^5$ CRPs. While, for this small example device, a model-based attack may be possible, a more realistic device for application with respect to this work would be comparable in size to that shown in \cite{Smith20}; generating at least $6.85 \times 10^{35}$ CRPs, far beyond a number feasibly tractable in numerical model-based attacks.

\section{Optical PUF Authentication}\label{sec:optical_puf_hw_auth}

PUFs have been suggested as a means to securely authenticate a networked device or remote user \cite{gassend2002controlled}. Current state-of-the-art means of authentication begins with the usage of a classical encryption key or token stored within a read-only memory (ROM) \cite{kim2018physical,pang2017optimization}. A PUF is of particular interest since they often form the basis of hardware primitives necessary to replace these cloneable shared encryption keys with a non-reproducible physical object or device.

\begin{figure}[htbp]
    \centering
    \includegraphics[width=\columnwidth,keepaspectratio]{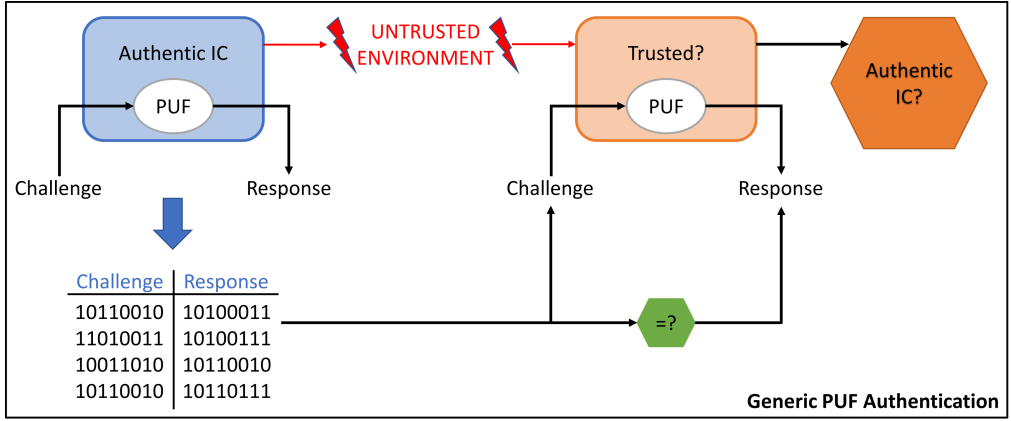}
    \caption{\textbf{Generic PUF application} detailing the operation of a PUF within a network or device communicating through an untrusted environment where the final value can be queried by a third-party to verify that the communication taking place is genuine. Once verified, communication can continue in an untrusted channel.}
    \label{fig:general_puf}
\end{figure}

The general operation of a PUF authentication system can be summarised by the image shown in Figure~\ref{fig:general_puf}, where the device containing the optical PUF is characterized with a set of challenges and the measured responses are captured by the verifier; these are called challenge-response pairs (CRPs). When the device is manufactured it is characterized with all, or a subset of, possible challenges and the responses are recorded. The CRPs are then securely stored and delivered to the purchaser of the device, often called the enrollment data. The CRPs in this protocol are not publicly released.

Due to aging of the device as discussed in \cite{Smith20,Afzal99} the responses to a challenge may change with time. For initial ageing of the device, allowing the verifier a threshold to compensate for the error is the easiest solution but, since this is a cumulative physical process, it is extremely hard to predict how each device will respond over longer periods of time and unfortunately would require a rebuild of the CRP tables by the manufacturer. This would be akin to a standard periodic recalibration procedure.

Once the device is in use away from the manufacturing facility and/or connected via an untrusted channel, one of the challenges can be applied to verify that the expected response is generated. If the verification is successful, the authenticity of the device can be assumed.

\subsection{Implementation} \label{sec:implementation_puf}
To utilize the optical PUF in a practical application, a more subtle approach is necessary. The challenge applied to the optical PUF is in the form of electrical settings sent through a control-module, while classical light, or single photons are present at the input ports. When the challenge applied is in the form of classical light, the MZIs will configure the light to a certain output profile, depicted by Figure~\ref{fig:mzi_puf_profile}. The output profile is in terms of normalized relative intensity across the measured photodiodes, where a distinct histogram is formed for each provided set of challenges to the MZIs. 

Similarly, the purely-quantum variant of utilizing the optical PUF will result in a set of MZI phase modulation settings being sent to the device as a challenge via the same control-module mentioned previously. The result will be an arbitrary state output from the device where the PUF has applied an arbitrary unitary transformation, $U^{arb}$, to the photons entering the device. The mathematical representation of a PUF authenticator with $N$ input and output modes and $q$-many unentangled photons will be:
\begin{align}\label{eq:puf_photon_auth}
    U^{arb}_{q_i} &\ket{\psi^{challenge}}_{q_i} = \ket{\psi^{response}}_{q_i}\,,\\ \nonumber
    &\text{where}\,\, \ket{\psi^{response}}_{q_i} = \sum_{j=0}^{N}{\alpha_{q_i,j} \ket{j}}\,,
\end{align}
for a $q$-many qubit (photon) input state with indexable qubits at position $i$. The response will be in a superposition state with some coefficients $\alpha_{q_i,j}$ on output modes $\ket{j}$ due to the natural structure of the PUF device \cite{Smith20}. Since a single or multi-photon input, of an unknown state will enter the device, the resulting output density matrix, $\rho_{\psi^{response}}$, will have an additional weight constants $\omega_n$ and PUF weight constants $\omega_p$, described further in Section~\ref{subsec:pfu_class_quant_inf_auth}, where each weight constant affects the separate logical $\ket{\mathbf{0}}$ and $\ket{\mathbf{1}}$ components of a qubit. The output response (histogram) from the PUF will then be projected from a measurement (actually repeated measurements) of:
\begin{align}
    \rho_{q_i\psi^{response}} &= \ket{\psi^{response}}_{q_i}\bra{\psi^{response}}_{q_i}\label{eq:simple_dens_resp}\\
    &= \omega_n \omega_p U^{\dag arb}_{q_i} (a\ket{01}\bra{01} + b\ket{10}\bra{10})U^{arb}_{q_i}\,.\label{eq:lr_dens_resp}%\\
%    &= \omega_n \omega_p U^{\dag arb}_{q_i} (c\ket{\,\mathbf{0}\,}\bra{\,\mathbf{0}\,} + d\ket{\,\mathbf{1}\,}\bra{\,\mathbf{1}\,})U^{arb}_{q_i}\,,\\ \nonumber
%    &\text{where}\,\, \ket{\mathbf{1}}=\frac{i}{\sqrt{2}} (\ket{\boldsymbol+} - \ket{\boldsymbol-})\,,
\end{align}
Where $q_i$ is each input qubit, the $\ket{01}$ and $\ket{10}$ are Fock states encoded in the $i$ and $i+1$ wave-guide input modes. Here we assume a separable qubit input state for clarity, however more complicated states such as entangled states and non-qubit states such as NOON states ($\ket{20}+\ket{02}$) can be applied.
%for differing polarization, either left = $\ket{\boldsymbol-}$, right = $\ket{\boldsymbol+}$, horizontal = $\ket{\,\mathbf{0}\,}$, or vertical = $\ket{\,\mathbf{1}\,}$.

\subsection{Classical PUF Readout}

The operation of a fully optical PUF in a quantum system has not been studied before but, something similar was approached by B.~{\v{S}}kori{\'c}, whereby a quantum readout protocol was developed to interface with a classical PUF \cite{vskoric2012quantum}. The readout protocol is modified, described below, with key notational differences to fit this work and to allow for a reconfigurable, optical, PUF.

%The quantum readout of the optical PUF is simple, where the challenge space of the device is a $d$-dimensional Hilbert space, $\mathcal{H}$, with a direct mapping to the response Hilbert space\footnote{Our device, being electronically reconfigurable, facilitates the mapping of an optical input combined with input phase settings into a `challenge' Hilbert space with the response from the device being a `response' Hilbert space.}
The quantum readout of the optical PUF is straightforward, where the challenge space of the device is a $d$-dimensional Hilbert space, $\mathcal{H}$, representing any state with $q$-qubits in $N$ input modes; with a projective mapping to the response Hilbert space. Our device, being electronically reconfigurable, facilitates the mapping of an optical input combined with input phase settings into a `challenge' Hilbert space with the response from the device being a `response' Hilbert space. An arbitrary input challenge $\ket{\psi^{challenge}} \in \mathcal{H}$ is mapped through the optical PUF, with statistical response $\hat{R}$, such that $\hat{R}\ket{\psi^{challenge}} \in \mathcal{H}$. The response will be unique, up to the limit of uniqueness from \cite{Smith20} where all nominal values of unique CRPs are above 50\%, for each challenge applied. In general, uniqueness should be approximately 50\%; this value is based on a binary PUF delivering results from $GF(2^n)$, where the reconfigurable optical device in question will show many more possibilities up to $CRP_{max}$ depending on the initial challenge, thus more unique and semi-unique CRPs exist. $\hat{R}$ may not necessarily be unitary, but can be decomposed into a response coefficient matrix $R$ and response unitary $U^{arb}$ such that $\hat{R}=R\,U^{arb}$. Here we note that our proposed protocol is a simple protective operation and does not require full state tomography; often prohibitively slow and difficult to accomplish.

The authentication between two parties, Alice and Bob, works where the verifier (Alice) wants to check if Bob still possesses the optical PUF. Alice first retrieves the original shared enrollment data, then picks a random state, $\psi$, and prepares the state $\psi^{challenge} \in \mathcal{H}$ and sends to Bob. Bob then lets the prepared particle interact with the optical PUF, resulting in the final response state $\psi^{response}=\hat{R}\psi^{challenge}$ that is then sent back to Alice. Since the result of the PUF response is in the density matrix, Alice then computes $\rho_{\psi^{response}}$ according to Equation~\ref{eq:lr_dens_resp}. Alice is then able to repeat this process multiple times to be sure that Bob's PUF is the correct PUF being used.

\subsection{Thresholding of Responses and Proportional Weights}
From Section~\ref{sec:implementation_puf}, a series of weights can be seen affecting the final output response. What the verifier sees as a response may vary with the weights based on the physical attributes of the underlying PUF. Ultimately a threshold on response would change the window of proportionality of the response such that the input selection weight constant for the device $\omega_n$ and the chosen PUFs CRP weight $\omega_p$ work together to make the overall weight of response. Since the device proposed to be used in this work is computer-controlled and interference-based, a threshold on response is required to get the verifier's window of acceptability. The probabilistic processes that influence the device's response necessitate an acceptability window on the response; used to make a determination on the response and whether to accept or reject the data. 

The form of the response from the readout will be a state composed of challenge-specific probabilities that can be verified against the original PUF being used for the message authentication. The constant on a single-response state can be represented by the general form $c\hat{R} U^{arb} \pm \epsilon$ where $c$ is a constant probability scalar and $\epsilon$ is the limiting threshold placed by the verifier when matching against the originally published CRPs.

\subsection{General Security Measure of Optical PUF Authentication}

The security of the protocol described is based on the no-cloning theorem, or in this work, the unclonability of the unknown quantum state by an eavesdropper \cite{dieks1982communication, bruss1999optimal}. For each round of the optical PUF quantum authentication protocol described: A standard challenge-estimation attack, where an adversary who attempts to determine the challenge applied using measurement techniques, will only have a maximum probability of $\tfrac{2}{(1+d)}$ to cause a `true' response from the PUF. The overall probability of a false positive decreases exponentially with the number of verification challenges that Alice sends to Bob. Since the protocol can be generalized to be $q$ qubits (photons), the state-space becomes $\ket{\psi^{challenge}}^{\tens{}q}$, where the attacker's per-qubit success probability is upper-bounded by $\tfrac{q+1}{q+d}$ \cite{vskoric2016security}.

\section{Authentication of Classical and Quantum Information with Fully-Optical PUF}\label{subsec:pfu_class_quant_inf_auth}

Traditionally, PUFs are only used for device authentication and not for message authentication. Our device is fully optical and can accept quantum states `at-once' unlike the original readout protocol \cite{vskoric2012quantum}. The modified protocol described also has the additional benefit of being reconfigurable, or the ability to act as many PUFs within a single device due to the tunability of the MZI's relative phases. The major difference is in the density matrix and projected measurements showing the additional $\omega_p$ parameter. The value for $\omega_p$ changes with each distinct challenge possible within the device such that the solution space increases not only by $\ket{\psi^{challenge}}^{\tens{}q}$ but, by an additional factor of: 
\begin{equation}
    \sum\limits_{i=0}^{CRP_{max}} \norm{\omega_{p,\,i}}\,,
\end{equation}
where the value for $CRP_{max}$ can be determined by a maximal upper bound by following the Catalan numbers shown in previous work \cite{Smith20}.

Following the modified quantum readout, a simple extension can be made where authentication of classical data, and quantum messages can be completed. In the classical case, Bob receives the optical PUF used in this work that acts as set of several distinct devices labeled $\{1,\ldots,CRP_{max}\}$. Bob wants to send an authenticated classical random variable $X:\Omega \to \mathbb{R}$ or message vector $\mathbf{X}=(x_1,\ldots,x_n)^T \in \mathbb{R}$ to Alice. Bob's message vector of length $n$ is decomposed into $k_i=\{1,\ldots,n\} \mapsto \mathbf{X}$, where $\mathbf{k} = (p_j)^N_{j=1} |_{p_j \in \{1,\ldots,CRP_{max}\}}$, for some PUF $p_j$ with selected CRP, $j$, and is subsequently sent to Alice. Effectively each element of a message vector, or random variable, maps to a specific PUF within the reconfigurable PUF, with a probability based on the message.

Alice and Bob perform the following:
\begin{enumerate}
    \item Bob sends $k_i$ to Alice over a public and non-authenticated channel.
    \item For $j\in \{1,\ldots,CRP_{max}\}$ Alice and Bob both perform the modified quantum readout protocol using the PUF's CRP number $p_j$.
\end{enumerate}
During each of the CRP tests in $p_j$, Alice slowly gains confidence that Bob's PUF is returning the responses to her issued challenges. Since there is a response to the challenges it can be assumed that the holder of Bob's PUF agrees with the individual variable $k_i$ sent over the non-authenticated channel. 

The quantum message authentication variant of the modified quantum readout protocol operates significantly differently with respect to the initial design. Consider a PUF design where $CRP_{max}=3$. Alice sends a random challenge state $\ket{\psi^{challenge}}$ to Bob. Bob then routes the challenge to $CRP_1$ with probability amplitude $\alpha$, $CRP_2$ with probability amplitude $\beta$, and $CRP_3$ with probability amplitude $\gamma$. The probability amplitudes are sent to satisfy $\abs{\alpha}^2 + \abs{\beta}^2 + \abs{\gamma}^2 = 1$ since the total probabilities cannot sum to be greater than one. Similarly, for more complex messages, a probability vector $\mathbf{P} = (p_i,\ldots,n)$, the probabilities must adhere to $\sum_{k=1}^{n} p_k^2 = 1$. This also assumes that $\norm{\mathbf{X}}_n = \norm{\mathbf{P}}_n$. Bob's response state sent back would then be:
\begin{align}
    \ket{\psi^{response}} &= \underbrace{\alpha \hat{R}_0}_{\alpha R_0 U_0^{arb}} \ket{\psi^{challenge}} + \underbrace{\beta \hat{R}_1}_{\beta R_1 U_1^{arb}} \ket{\psi^{challenge}} + \underbrace{\gamma \hat{R}_2}_{\gamma R_2 U_2^{arb}} \ket{\psi^{challenge}}\,,
\end{align}
which is subsequently sent to Alice. Alice is then able to verify, even though she doesn't know the probability amplitudes $(\alpha,\beta,\gamma)$, since she does know the components of $\hat{R}_i$ from the initial registration of the optical PUF. This means that when Alice verifies Bob's response, that she will need to rely on the initial PUF weight constant, $\omega_p$, to have a `best guess' of what the probability assignment for the PUF's CRPs would be when assigned by Bob. Alice then knows that:
\begin{equation}
    \abs{\omega_p^{CRP_0}}^2 + \abs{\omega_p^{CRP_1}}^2 + \abs{\omega_p^{CRP_2}}^2 \propto \abs{\alpha}^2 + \abs{\beta}^2 + \abs{\gamma}^2\,,
\end{equation}
where she will then be able to determine that the responses and probabilities match those that were originally registered from the PUF -- assumed to be -- held by Bob. Alice also knows from receiving the modified state that the sender \textit{has} to be holding Bob's PUF; thus achieving an optical, reconfigurable, PUF-based authentication of a quantum state. 

\subsection{Secrecy of Modified Readout for Reconfigurable PUF}
From a security standpoint of the quantum message authentication using an all-optical PUF with reconfigurable CRPs, the data could still be considered confidential. Assuming a challenge-estimation attack on a $q$-qubit system where $q < d$, an initial state $\psi^{challenge}$ would be chosen uniformly at random. An attacker could know $\psi^{response}$ but would not posses the PUF. Additionally assuming the attacker does not have access to a quantum machine, or any device that can compute arbitrary unitary transformations losslessly, only a generic measurement could be completed with a biased estimator. The adversary would then only be able to compute an estimation of a response $\hat{\mathbb{E}}_{\ket{\psi^{response}}} = \hat{R}\hat{\mathbb{E}}_{\ket{\psi^{challenge}}}$.

 If an adversary challenged Bob's PUF, the response would not necessarily reveal the probability amplitudes $(\alpha,\beta,\gamma)$ of the proper CRP because, to an adversary, this information could plausibly be from a different reconfigurable PUF or could relate to a different reconfiguration setting. In addition, an adversary that could determine the probability amplitudes sent for different CRPs would not know the original registered parameter, $\omega_p$, that contains the true suggested probability amplitudes for each of the CRPs within the reconfigurable PUF. 

Thus, the modified quantum readout scheme for a reconfigurable all optical PUF verifies the authenticity of the PUF and can be used to authenticate both classical messages and quantum states.

\section{Conclusion}
In this work we described the usage of a reconfigurable all-optical PUF as a hardware authentication mechanism. The usage of the optical PUF as an authentication mechanism is carried out by a modified quantum readout protocol. The readout protocol makes the PUF able to authenticate classical and quantum information through the usage of classical light or single-photon-level manipulations. Additionally, the usage of the hardware optical PUF enables one to authenticate that the receiver of information is not adversarial in nature. This work represents the first application of an all-optical reconfigurable PUF for tasks other than object and direct-access user authentication.

\medskip

\noindent\textbf{Acknowledgements.} 
H~S.~Jacinto would like to thank AFRL for fellowship support. The authors would like to thank J.~E.~Schneeloch for helpful insight and discussions. Any opinions, findings, conclusions, or recommendations expressed in this material are those of the authors and do not necessarily reflect the views or endorsement of the Air Force Research Lab.

\medskip

\noindent\textbf{Disclosures.} The authors declare no conflicts of interest.

%\section{References}

\bigskip

% Bibliography
\bibliography{references}

% Full bibliography added automatically for Optics Letters submissions; the following line will simply be ignored if submitting to other journals.
% Note that this extra page will not count against page length
%\bibliographyfullrefs{references}

% Please include bios and photos of all authors for aop articles
%\ifthenelse{\equal{\journalref}{aop}}{%
%\section*{Author Biographies}
%\begingroup
%\setlength\intextsep{0pt}
%\begin{minipage}[t][6.3cm][t]{1.0\textwidth} % Adjust height [6.3cm] as required for separation of bio photos.
%  \begin{wrapfigure}{L}{0.25\textwidth}
%    \includegraphics[width=0.25\textwidth]{john_smith.eps}
%  \end{wrapfigure}
%  \noindent
%  {\bfseries John Smith} received his BSc (Mathematics) in 2000 from The University of Maryland. His research interests %include lasers and optics.
%\end{minipage}
%\begin{minipage}{1.0\textwidth}
%  \begin{wrapfigure}{L}{0.25\textwidth}
%    \includegraphics[width=0.25\textwidth]{alice_smith.eps}
%  \end{wrapfigure}
%  \noindent
%  {\bfseries Alice Smith} also received her BSc (Mathematics) in 2000 from The University of Maryland. Her research %interests also include lasers and optics.
%\end{minipage}
%\endgroup
%}{}

\end{document}